%% file: my_MSC_paper_ICASSP.tex
\documentclass{article}
\usepackage{spconf,mathtools}
\usepackage{amssymb}
\usepackage{dsfont}
\usepackage{multirow}
\usepackage{graphicx}
\usepackage[table]{xcolor}
\usepackage{algorithm}
\usepackage{algpseudocode}
\usepackage{color}

\usepackage{mathrsfs}
\DeclareSymbolFontAlphabet{\mathrsfs}{rsfs}
\usepackage[mathscr]{eucal}

\usepackage{amsbsy}

\usepackage{bm}

\include{macrofile2}

\begin{document}
\title{Multilinear Subspace Clustering}

\name{Eric Kernfeld$^\ast$, Nathan Majumder$^\dagger$, Shuchin Aeron$^\dagger$, and Misha Kilmer$^\dagger$}
\address{$^\ast$ University of Washington, Seattle, WA, USA. $^\dagger$ Tufts University, Medford, MA, USA. \thanks{This work is part of the first author's senior honors thesis at Tufts University. Nathan Majumder, Shuchin Aeron and Misha Kilmer were supported by NSF grant 1319653.}}
\vspace{-10mm}
%


\maketitle
\begin{abstract}
In this paper we present a new model and an algorithm for unsupervised clustering of 2-D data such as images. We assume that the data comes from a union of multilinear subspaces (UOMS) model, which is a specific structured case of the much studied union of subspaces (UOS) model. For segmentation under this model, we develop Multilinear Subspace Clustering (MSC) algorithm and evaluate its performance on the YaleB and Olivietti image data sets. We show that MSC is highly competitive with existing algorithms employing the UOS model in terms of clustering performance while enjoying improvement in computational complexity. 

\end{abstract}

\vspace{-3mm}
\section{Introduction}

Most clustering algorithms seek to detect disjoint clouds of data. However, in high-dimensional statistics, data can become very sparse, and these types of methods have trouble dealing with noise. In fact, a completely new approach to the geometry of clustering has recently made headway in the analysis of high-dimensional data sets. Called subspace clustering, this approach assumes that data come from subspaces offset at angles, rather than from clouds offset by gaps, the so called Union of Subspaces (UOS) model \cite{Vidal1,tsc,Park2014_GSC}. Applications have included detection of tightly correlated gene clusters in genomics \cite{KiWi}, patient-specific seizure detection from EEG data \cite{DuttaEEGSC}, and image segmentation \cite{Yang:EECS-2006-195}. 

All subspace clustering methods must embed data in $\mathbb{R}^n$. However, in some of the high-dimensional data sets where subspace clustering has been applied, the initial structure of the data is not a vector but rather a matrix or tensor (multi-way array). Examples include the auditory temporal modulation features in \cite{Panagakis:2014kx}, the image patches in \cite{Yang:EECS-2006-195}, and raw EEG data under the ``sliding-window approach'' \cite{DuttaEEGSC}. We seek to develop a clustering method that incorporates the geometric innovation of subspace clustering without vectorizing these higher-order arrays. To do this, we formulate an algebraic generative model for the data, along with methods for inference. 



\textbf{The Subspace Clustering Problem and a Multilinear Variant} - Mathematically, the subspace clustering problem is described as follows. Given a set of points ${x}_n$, $n=1...N$, suppose each point is an element of one of the $K$ subspaces. The problem is to decide membership for each of the $N$ points. For simplicity, we treat $K$ as known. 

In order to take advantage of patterns in two-way data, we modify the assumptions of the subspace clustering problem. Rather than modeling the data as a union of subspaces, we assume they come from a union of \emph{tensor products} of subspaces \cite{Hackbusch_book}. Given subspaces $\mathcal{U} \subset \mathbb{R}^n$ and $\mathcal{V} \subset \mathbb{R}^m$, suppose the columns of $\M{U}$ form a basis of $\mc{U}$ and likewise for $\M{V}$ and $\mathcal{V}$. The tensor product $\mathcal{U}  \otimes \mathcal{V}$ is the set $\{\M{A}|\M{A}=\M{UYV}^T\}$, where $\M{Y}$ is a dim($\mc{U}$)$\times$ dim($\mc{V}$) matrix. In other words, this is a set of matrices with (column/row) space confined to ($\mathcal{U}$/$\mathcal{V}$). We refer to this model as the union of multilinear subspaces (UOMS) model and we call this the multilinear subspace clustering (MSC) problem. \textbf{\emph{Note that while $\mathcal{U}  \otimes \mathcal{V}$ is a \emph{tensor-subspace} of the tensor space $\mathbb{R}^{n} \otimes \mathbb{R}^{m}$, not all subspaces of the tensor space is a tensor subspace \cite{Hackbusch_book}. Therefore we are assuming a tensor-subspace structure on the clusters under the UOMS model.}}

The difference between the generative models for UOS and UOMS is clarified in Algorithms \ref{alg:UOS}, \ref{alg:UOMS}.

\begin{algorithm}
\caption{UOS Data Generation: $N$ points, $K$ clusters of latent dimension $d$ and ambient dimension $D$}
\label{alg:UOS}
Given $\{\M{U}_1, ..., \M{U}_K\} \in \mathbb{R}^{D\times d}$,\\
Repeat $N$ times:\\
Draw $k$ from $\{1, ..., K\}$ \\
Draw a random length-$d$ vector $y_n$\\
Compute datum $\V{x}_n=\M{U}_k\V{y}_n$
\end{algorithm}
\vspace{-5mm}

\begin{algorithm}
\caption{UOMS Data Generation: $N$ points, $K$ clusters of latent dimension $d_v d_u$ and ambient dimension $D_v D_u$}
\label{alg:UOMS}
Given $\{\M{U}_1, ..., \M{U}_K\} \in \mathbb{R}^{D_u\times d_u}, \{\M{V}_1, ... , \M{V}_K\} \in \mathbb{R}^{D_v\times d_v}$\\
Repeat $N$ times:\\
Draw $k$ from $\{1, ..., K\}$ \\
Draw a random $d_u$ by $d_v$ matrix $\M{Y}_n$\\
Compute datum $\M{A}_n=\M{U}_k\M{Y}_n\M{V}_k^T$
\end{algorithm}
\vspace{-3mm}

\textbf{Relation of UOMS to existing subspace models} - Note that the UOS model with single subspace (one cluster) is related to the Principal Component Analysis (PCA). Similarly the UOMS with one cluster is closely related to separable covariance models \cite{hoff2011} and also 2D-PCA\cite{2DPCA}. Further in \cite{hoff2011}, extensions of this idea to 3-D, 4-D,... data is shown to be equivalent to HOSVD and Tucker decompositions \cite{Kolda:2009vn} that have been useful for dimensionality reduction for image ensembles \cite{Vasilescu_2003}. Further such multilinear subspace models have been used in machine learning \cite{HeCN05_NIPS,MPCA_2006}. In this paper we study an extension of these models by considering a \emph{union} of such structured subspaces. 

\vspace{0pt}  
\begin{algorithm}
\caption{Thresholded Subspace Clustering (TSC)}
\label{tscAlg}
\begin{algorithmic}
\item \textbf{Input}
	\State $\M{X} \in \mathbb{R}^{D \times N}$ holding data vectors $\V{x}_i$, number of clusters $K$, thershold ($1 \le q \le N$)
\item \textbf{Procedure}
	\State Normalize $\V{x}_i$, Compute adjacency matrix $C = |X^\top X|$
	\State Set all but q highest values of each row of C to zero
	\State $\M{C}_{i,j} = \exp\left(-2*\cos^{-1}(\M{C}_{ij})\right)$
	\State Perform normalized spectral clustering on $\M{C}$
\item \textbf{Output}
	\State A vector in $\mathbb{R}^N$ with clustering labels for all $\V{x}_i$
\end{algorithmic}
\end{algorithm}
\vspace{-5mm}
\vspace{-5pt}  
\begin{algorithm}
\caption{Sparse Subspace Clustering (SSC)}
\label{sscAlg}
\begin{algorithmic}
\item \textbf{Input}
	\State $\M{X} \in \mathbb{R}^{D \times N}$ holding data vectors $\V{x}_i$, $\lambda_1, \lambda_2 \geq 0$
	\State the number of clusters $K$
\item \textbf{Procedure}
	\State Solve $\arg\min_{\M{C},\M{E}} \| \M{X} - \M{X}\M{C} - \M{E}\|_{F}^{2} + \lambda_1 ||\M{C}||_1 + \lambda_2\|\M{E}\|_{1}$ s.t. $\M{C}_{ii} = 0$
	\State Form $\M{W} = |\M{C}| + |\M{C}|^\top$
	\State Apply spectral clustering on $\M{W}$
\item \textbf{Output}
	\State A vector in $\mathbb{R}^N$ with clustering labels for all $\V{x}_i$
\end{algorithmic}
\end{algorithm}
\textbf{Clustering under the UOS model} - There are many algorithms exploiting the UOS model for clustering, but we focus on two general methods that form an affinity matrix among data points followed by spectral clustering \cite{Luxburg:2007dq}. The first, called Thresholded Subspace Clustering (TSC), is introduced in \cite{tsc}. This provably reliable and robust subspace clustering algorithm involves constructing a weighted graph where nodes represent the data points and edges represent the connectivity of any two points. The inner product between the data points is used as the edge weight with the idea that points in the same subspace will generally have a higher inner product than points in different subspaces. The symmetric adjacency matrix for the graph is then thresholded, setting all but the $q$ highest weights in each row to zero, in order to filter out noise.
The second method, called Sparse Subspace Clustering (SSC) \cite{ssc} involves expressing each point as a linear combination of the other points in the dataset. The algorithm finds the sparsest possible linear representation of each data point in terms of other data points -- achieved by minimizing the $\ell_{1}$-norm -- with the idea that the points used will come from the same subspace as the point in question. A weighted graph is then formed with an adjacency matrix found using the sparse representations of each point. Both the TSC and SSC algorithms, taken from \cite{tsc} and \cite{ssc} respectively, are detailed in Algorithms \ref{tscAlg} and \ref{sscAlg}. 

\vspace{-5mm}

\section{Clustering under the UOMS model}
\vspace{-3mm}
In the case of two-way data, our data points would be a collection of $N$ matrices $\M{A}_n \in \mathbb{R}^{D_U \times D_V}$ such that the columns come from a union of subspaces $\mathcal{U}_1 \cup \hdots \cup \mathcal{U}_K$ and the rows come from a union of subspaces $\mathcal{V}_1 \cup \hdots \cup \mathcal{V}_K$. To take advantage of this fact and find these $\mathcal{U}_i$ and $\mathcal{V}_i$ subspaces, one method would be to cluster all $D_{\mathcal{V}}N$ columns and all $D_{\mathcal{U}}N$ rows separately; however, this is an expensive solution. Instead, we randomly select a single column and a single row from each matrix and cluster these. We stack the random columns side by side to form a $D_{\mathcal{U}} \times N$ matrix $X_{cols}$ and transpose and stack our random rows side by side to form a $D_{\mathcal{V}} \times N$ matrix $\M{X}_{rows}$. The $i$th column of each of these matrices comes from the $i$th ($i=1,...,N$) data matrix $\M{A}_i$. We then perform a clustering algorithm on $\M{X}_{rows}$ and $\M{X}_{cols}$ separately, but pause after obtaining the symmetric adjacency matrix C in each case. We repeat this process for $T$ trials, ending up with $2T$ adjacency matrices, which we then combine in one of a few possible ways.  

Possible combination methods are detailed subsequetly. Combining these adjacency matrices can be thought of as condensing multiple graph realizations to obtain a single weighted graph representing our clustering problem. Once we have our condensed graph, we perform spectral clustering on it to achieve the segmentation of our original points. Algorithm \ref{mscAlg} outlines the steps described above.

\vspace{-2mm}
\begin{algorithm}
\caption{Multilinear Subspace Clustering (MSC)}
\label{mscAlg}
\begin{algorithmic}
\item \textbf{Input:}
	\State Data $\M{A}_1,...,\M{A}_N \in \mathbb{R}^{D_c \times D_r}$, number of clusters $K$
	\State Clustering Method (TSC or SSC), number of trials $T$
\item \textbf{Procedure:}
	\State \textbf{For} $T$ trials
	\State \hspace{\algorithmicindent} form $\M{X}_{cols} \in \mathbb{R}^{D_c \times N}$ with $i$-th column randomly selected from $\M{A}_i$
	\State \hspace{\algorithmicindent} form $\M{X}_{rows} \in \mathbb{R}^{D_r \times N}$ with $i$-th column randomly selected from $\M{A}_i^\top$
	\State \hspace{\algorithmicindent} Run clustering (TSC or SSC) on $\M{X}_{cols}$ and $\M{X}_{rows}$ to get $\M{C}_{cols} \in \mathbb{R}^{N \times N}$ and $\M{C}_{rows} \in \mathbb{R}^{N \times N}$
	\State \textbf{End For}
	\State Combine all $\M{C}_{cols}$ and $\M{C}_{rows}$ into a single adjacency matrix $\M{C} \in \mathbb{R}^{N \times N}$
	\State Perform spectral clustering on $\M{C}$
\item \textbf{Output:}
	\State A vector in $\mathbb{R}^N$ with clustering labels for all $\M{A}_i$
\end{algorithmic}
\end{algorithm}
\textbf{Combining the Graph Realizations} - We now discuss several (heuristic) methods for combining the adjacency matrices obtained at each iteration of MSC.
\begin{enumerate}
\setlength\itemsep{-3pt}
\item \textbf{Addition}: One simple method is to add the $2T$ adjacency matrices together.

\item \textbf{Thresholding}: Add the matrices together followed by thresholding, setting all but the $q$ highest edges per row to zero. A possible choice of threshold for this method would be the average number of data points per cluster -- if this number is known -- minus one (to count out the point itself).

\item \textbf{Filtering by Quantile}: A "quantile" method involves choosing a parameter $l$ and taking the $l$-th highest weight at each edge out of all the adjacency matrices. The choice of $l$ poses an obstacle, as there is not a given value that will be optimal for all graphs. 

\item \textbf{Projection}: Project each individual adjacency matrix's columns onto its leading $K$ singular vectors (corresponding to largest singular values) before adding the instances.  However, the fact that each matrix is projected onto its leading singular vectors before sharing any information with other graph realizations could lead to loss of quality.
\end{enumerate}

\textbf{Remark} - These methods are by no means exhaustive. In particular, the problem of combining various graph realizations for the same problem instance by itself is an interesting avenue of research.

\textbf{Algorithmic Complexity} -
For $N$ data points of dimension $D$, TSC has algorithmic complexity $\mc{O}(DN^2)$. Therefore, if we are comparing TSC on vectorized 2-way data against MSC using TSC on the same data in matrix form, the MSC data points will be matrices of size $D_c \times D_r$ where $D=D_cD_r$. At each iteration of MSC, we form the matrices of size $D_c \times N$ and $D_r \times N$ for the column space and row space respectively. Since we then perform TSC on these matrices, the algorithmic complexity at each iteration will be $\mathcal{O}(D_cN^2 + D_rN^2)$ or approximately $\mc{O}(\sqrt{D}N^2)$ when $D_c \approx D_r$. For the projection method, which is the computationally most expensive, when $K << N$, using randomized SVD a computational cost of ${\cal O}(N^2 \log K)$ \cite{Halko_2011} is incurred. Therefore, for T iterations of MSC, we have $\mc{O}(T\sqrt{D}N^2)$ compared to $\mc{O}(DN^2)$ for TSC. Therefore if we can pick a number of trials T such that $T \ll \sqrt{D}$, MSC will be cheaper. This obviously leads to a possible conclusion that MSC will be a better choice for large data while TSC will be more realistic for data of smaller dimensions. The computational complexity of the SSC algorithm is ${\cal O}(DN^3)$, which for large $D$ and $N$ becomes more prohibitive compared to TSC as well as MSC. 

\vspace{-3mm}
\section{Numerical Results}
\subsection{Yale Face Database}

We first test our Multilinear Subspace Clustering algorithm on images of faces with fixed pose and expression under different illumination conditions from the Yale Face Database B. These images are of size $192 \times 168$ pixels and we use 64 images each of 37 different people. The parameters that we vary are the number of trials $T$ and the method of combining graph realizations. Over numbers of clusters ranging from 2 to 10, we compare results to those obtained in \cite{tsc} and \cite{ssc} using TSC and SSC. We measure success based on a clustering error that measures the fraction of misclassified points and is defined in \cite{tsc}.

\subsection{Results of Different Methods for Combining Graphs}
We first test the various methods of condensing our graph that we detailed in section 2. We set $T = 15$ as a constant number of trials and find the clustering error for 2, 5, and 10 clusters using all four methods. The results, depicted in Table 1, show that the method of projecting the columns of the adjacency matrices onto their leading eigenvectors before adding them produces the smallest clustering error by far. Therefore we use this method for future results shown in this paper.

\begin{table}
\centering
\caption{The clustering error for our four methods for condensing the graph obtained in MSC.}
\label{tab:1}
\begin{tabular}{|c|c|c|c|c|}
\hline
\multicolumn{2}{|c|}{\multirow{2}{*}{\% MSC Clustering Errors}}  & \multicolumn{3}{c|}{Number of Clusters} \\ \cline{3-5} 
\multicolumn{2}{|c|}{}                                    & 2           & 5           & 10          \\ \hline
\multirow{4}{*}{Method} & addition   & 12.21      & 44.16      & 48.96      \\ \cline{2-5} 
                                             & projection & 4.06      & 20.19      & 38.70      \\ \cline{2-5} 
                                             & quantile   & 20.75      & 47.37      & 45.52      \\ \cline{2-5} 
                                             & threshold  & 11.13      & 35.52      & 46.15      \\ \hline
\end{tabular}
\end{table}

\subsection{Varying the Number of Trials}
Using the project-first method of combining our graph realizations, we next test the effect of varying the number of trials used for MSC. The goal is to find a number independent of the number of data points that is small enough such that MSC remains as efficient as possible (based on the discussion on computational complexity) while minimizing the clustering error. From Table \ref{tab:2} We note that the clustering error begins to plateau at 100 - 200 trials. We are randomly selecting columns and rows from each matrix, so ideally we get a different fiber at each iteration. Setting the number of trials to be more than the number of fibers does not make sense if we do not want to use a fiber more than once. In the following subsection, when we discuss MSC using SSC, we use 100 trials, as it seems to be where increasing the trials begins to have less effect on performance.


\begin{table}
\centering
\caption{The MSC using TSC clustering error data}
\label{tab:2}
\label{varyNumTrialsTab}
\begin{tabular}{|c|c|c|c|c|}
\hline
\multicolumn{2}{|c|}{\multirow{2}{*}{\% MSC Clustering Errors}{\begin{tabular}[c]{@{}c@{}}\\ \end{tabular}}} & \multicolumn{3}{c|}{Number of Clusters} \\ \cline{3-5} 
\multicolumn{2}{|c|}{}                                                                                                       & 2           & 5           & 10          \\ \hline
\multirow{7}{*}{Number of Trials}                          & 10					   & 5.50      & 22.61	 & 43.85	    \\ \cline{2-5}
									  & 20                                            & 4.44      & 19.01      & 40.73      \\ \cline{2-5} 
                                                                             & 50                                            & 3.37      & 15.68      & 38.28      \\ \cline{2-5} 
                                                                             & 100                                           & 2.65      & 14.43      & 36.82      \\ \cline{2-5} 
                                                                             & 200                                           & 2.32      & 13.73      & 34.95      \\ \cline{2-5} 
                                                                             \hline
\end{tabular}
\end{table}

\subsection{Varying the SSC Parameters}
The SSC code taken from \cite{ssc} requires parameter that were previously optimized for vectorized data. We therefore vary three of these inputs to achieve the best results. The first two parameters allow for the algorithm to reject outliers and allow for affine subspaces respectively. If the outlier parameter is "true", the SSC code will dismiss outlying entries, and if the affine parameter is "true", the code will broaden its scope to include affine subspaces. The numerical results can be seen in Table \ref{MSCwSSC:outlier_affine}. Interestingly, the best results come from allowing outlier rejection but disallowing affine subspaces. 

\textbf{Note}: This is slightly counterintuitive for MSC, as SSC is being run on a matrix of single fibers, not on an entire data point at each iteration. Therefore, one might think that since many of the fibers will have no connection to each other, random fibers would get thrown out as outliers, when really we would want to get rid of entire outlying data points, not single fibers. This is an important avenue for future research. 

\begin{table}
\centering
\caption{MSC using SSC}
\label{MSCwSSC:outlier_affine}
\begin{tabular}{|c|c|c|c|c|c|}
\hline
\multicolumn{3}{|c|}{\multirow{2}{*}{\% MSC CE}{\begin{tabular}[c]{@{}c@{}}\end{tabular}}}                             & \multicolumn{3}{c|}{Number of Clusters} \\ \cline{4-6} 
\multicolumn{3}{|c|}{}                                                                                                                       & 2           & 5           & 10          \\ \hline
\multirow{4}{*}{Constraints} & \begin{tabular}[c]{@{}c@{}}Outliers\\ ON\end{tabular}  & \begin{tabular}[c]{@{}c@{}}Affine\\ ON\end{tabular}  & 10.02      & 58.17      & 70.26      \\ \cline{2-6} 
                             & \begin{tabular}[c]{@{}c@{}}Outliers\\ ON\end{tabular}  & \begin{tabular}[c]{@{}c@{}}Affine\\ OFF\end{tabular} & 2.71      & 21.22      & 40.68      \\ \cline{2-6} 
                             & \begin{tabular}[c]{@{}c@{}}Outliers\\ OFF\end{tabular} & \begin{tabular}[c]{@{}c@{}}Affine\\ ON\end{tabular}  & 25.57      & 59.51      & 68.80      \\ \cline{2-6} 
                             & \begin{tabular}[c]{@{}c@{}}Outliers\\ OFF\end{tabular} & \begin{tabular}[c]{@{}c@{}}Affine\\ OFF\end{tabular} & 25.57      & 59.51      & 68.80      \\ \hline
\end{tabular}
\end{table}

%

\textbf{Overall Performance Comparisons} - We first compare the use of TSC on vectorized data to MSC using TSC at each iteration (on data in matrix form). The average clustering errors for TSC given in \cite{tsc} are (without preprocessing) 12.42\%, 29.17\%, 39.84\% for 2,5, and 10 clusters respectively. The full set of data from varying the number of MSC trials can be seen in Table 2, but for 100 trials, the number at which the decrease in error begins to plateau, we obtain average clustering errors of 2.65\%, 14.43\%, and 36.82\% for 100 trials for 2, 5, and 10 clusters respectively on the same data. Lowering the number of trials we use to 50 and even 20, we still obtain better clustering errors using MSC for low numbers of clusters and comparable clustering errors for 10 clusters.

Because each face datum is so large, we cannot compare the SSC algorithm to MSC using SSC. Vectorizing the data yields points of dimension $192*168=32256$, which is far too large to use the SSC algorithm from \cite{ssc}, given the optimization problem of constructing a sparse representation for each point. This therefore is an immediate advantage for MSC, as SSC is impractical for any very large data. We do report clustering errors for MSC using SSC at each iteration, as can be seen in the previous subsection. The optimized results are, for 2, 5, and 10 clusters respectively, 2.31\%, 15.43\%, and 36.67\%, which are comparable to those obtained from MSC using TSC on the same data. The results from each algorithm are shown in Table \ref{YaleClusteringErrorsTable}.

\begin{table}
\centering
\caption{The clustering errors for each algorithm on the Yale Faces dataset using 2, 5, and 10 clusters}
\label{YaleClusteringErrorsTable}
\begin{tabular}{|c|c|c|c|c|}
\hline
\multicolumn{2}{|c|}{\multirow{2}{*}{\% Clustering Errors}} & \multicolumn{3}{c|}{Number of Clusters} \\ \cline{3-5} 
\multicolumn{2}{|c|}{}                                      & 2           & 5           & 10          \\ \hline
\multirow{4}{*}{Algorithm}          & TSC                   & 12.42       & 29.17       & 39.84       \\ \cline{2-5} 
                                    & MSC using TSC         & 2.65        & 14.43       & 36.82       \\ \cline{2-5} 
                                    & SSC                   & \multicolumn{3}{c|}{timed out}      \\ \cline{2-5} 
                                    & MSC using SSC         & 2.31        & 15.43       & 36.67       \\ \hline
\end{tabular}
\end{table}

\vspace{-3mm}
\subsection{Olivetti Face Database}
We look at results of the algorithms on the Olivetti Faces data, found at \cite{olivetti}. It includes $400$ images of size  $64 \times 64$ pixels, $10$ each of $40$ different subjects. For each test, we perform and average results from $100$ runs, each of which chooses a random set of subjects and takes all $10$ images from each. The results are shown in Table \ref{OlivettiClusteringErrorsTable}.

\begin{table}
\centering
\caption{The clustering errors for each algorithm on the Olivetti Faces dataset using 2, 5, and 10 clusters}
\label{OlivettiClusteringErrorsTable}
\begin{tabular}{|c|c|c|c|c|}
\hline
\multicolumn{2}{|c|}{\multirow{2}{*}{\% Clustering Errors}} & \multicolumn{3}{c|}{\# Clusters} \\ \cline{3-5} 
\multicolumn{2}{|c|}{}                                        & 2           & 5           & 10          \\ \hline
\multirow{4}{*}{Algorithm}           & TSC                    & 7.70      & 23.78      & 25.01      \\ \cline{2-5} 
                                     & MSC using TSC          & 4.40      & 14.04      & 20.79      \\ \cline{2-5} 
                                     & SSC                    & 2.90      & 9.72      & 18.10      \\ \cline{2-5} 
                                     & MSC using SSC          & 2.30      & 10.70      & 16.28      \\ \hline
\end{tabular}
\end{table}
\vspace{-2mm}
\section{Conclusions and Future work}
In this paper we presented a new model, namely the UOMS model and presented an algorithm unsupervised clustering under this model. We showed that the resulting algorithm is competitive with existing methods while being computationally more efficient. For future work, we will investigate how to deal with the outliers when we are drawing rows and columns from the data and not the entire data itself. Another important avenue is to develop a systematic method with provable guarantees for combining various graph realizations. As is well known, subspace clustering performance depends on the distribution of the data. Therefore, if the relative importance of different instances can be characterized one may perform a weighted graph combination. We will also extend this method for automatic clustering of 3-D data sets such as action videos.

\newpage
\bibliographystyle{IEEEbib.bst}
\bibliography{my_MSC_paper,eric_kernfeld_biblio}

\end{document}

%% file: macrofile2.tex
\newcommand{\Sec}[1]{\hyperref[sec:#1]{\S\ref*{sec:#1}}} 
\newcommand{\Eqn}[1]{\hyperref[eq:#1]{(\ref*{eq:#1})}} 
\newcommand{\Fig}[1]{\hyperref[fig:#1]{Figure~\ref*{fig:#1}}} 
\newcommand{\Tab}[1]{\hyperref[tab:#1]{Table~\ref*{tab:#1}}} 
\newcommand{\Thm}[1]{\hyperref[thm:#1]{Theorem~\ref*{thm:#1}}} 
\newcommand{\Lem}[1]{\hyperref[lem:#1]{Lemma~\ref*{lem:#1}}} 
\newcommand{\Prop}[1]{\hyperref[prop:#1]{Property~\ref*{prop:#1}}} 
\newcommand{\Cor}[1]{\hyperref[cor:#1]{Corollary~\ref*{cor:#1}}} 
\newcommand{\Def}[1]{\hyperref[def:#1]{Definition~\ref*{def:#1}}} 
\newcommand{\Alg}[1]{\hyperref[alg:#1]{Algorithm~\ref*{alg:#1}}} 
\newcommand{\Ex}[1]{\hyperref[ex:#1]{Example~\ref*{ex:#1}}} 

\newcommand{\mc}[1]{\mathcal{#1}}

\newcommand{\V}[1]{{\bm{\mathbf{\MakeLowercase{#1}}}}} 

\newcommand{\M}[1]{{\bm{\mathbf{\MakeUppercase{#1}}}}} 

